\documentclass[12pt]{article}

\usepackage{amssymb,textcomp,amsmath,graphicx,color,bigints,multirow,subfig,hyperref}
\usepackage{caption}

\DeclareMathAlphabet{\mathcal}{OMS}{cmsy}{m}{n}
\long\def\symbolfootnote[#1]#2{\begingroup
\def\thefootnote{\fnsymbol{footnote}}
\footnote[#1]{#2}\endgroup}

\numberwithin{equation}{section}

\newcommand{\ie}{{\em i.e.~}}
\newcommand{\eg}{{\em e.g.~}}

\newcommand{\be}{\begin{equation}}  \newcommand{\ee}{\end{equation}}
\newcommand{\bea}{\begin{eqnarray}} \newcommand{\eea}{\end{eqnarray}}

\newcommand{\bse}{\begin{subequations}}
\newcommand{\ese}{\end{subequations}}
\newcommand{\bi}{\begin{itemize}}
\newcommand{\ei}{\end{itemize}}
\newcommand{\lp}{\left(}
\newcommand{\rp}{\right)}
\newcommand{\lb}{\left[}
\newcommand{\rb}{\right]}

\newcommand{\bh}{\rm PBH}

\begin{document}

\begin{center}
\Large{Hierarchical Merger of Primordial Black Holes in Dwarf Galaxies}
\end{center}
\begin{center}
\large{Encieh Erfani$^{1,\,*}$}, \large{Tadeo D.~Gomez-Aguilar$^{2,\,3,\,\dag}$} and \large{Juan Carlos Hidalgo$^{3,\,\ddag}$}
\end{center}
\begin{center}
\textit{$^1$Department of Physics, Institute for Advanced Studies in Basic Sciences (IASBS), Zanjan 45137-66731, Iran\\
$^2$Facultad de Ciencias en F\'{i}sica y Matem\'{a}ticas, Universidad Aut\'{o}noma de Chiapas, 29050, Tuxtla Guti\'{e}rrez, Chiapas, M\'{e}xico\\
$^3$Instituto de Ciencias F\'{i}sicas, Universidad Nacional Aut\'{o}noma de M\'{e}xico, 62210, Cuernavaca, Morelos, M\'{e}xico}

\end{center}

\date{}

\symbolfootnote[0]{$^{*}$erfani@iasbs.ac.ir}
\symbolfootnote[0]{$^{\dag}$tadeo.dga@icf.unam.mx}
\symbolfootnote[0]{$^{\ddag}$hidalgo@icf.unam.mx}

\begin{abstract}
We study the merger history of primordial black holes (PBHs) in a scenario where they represent the dominant dark matter component of a typical dwarf galaxies' core. We investigate the possibility of a sequence of collisions resulting in a hierarchical merger of black holes, and look at the final mass spectrum in such {\it clusters}, which initially present a monochromatic (single-mass) PBH population. Our study shows that the merging process results in the transfer of about $40\%$ of the total mass of the core to the merger products regardless of the initial mass of PBHs, with about $5\%$ of energy radiated out in the form of gravitational waves. We find that, in the lighter mass limit, black holes up to eight times more massive than the original population can be formed within a Hubble time. 
\end{abstract}

\newpage

\section{Introduction}
\paragraph{}

Observation of gravitational waves (GWs) within the last six years is transforming the field of astronomy and our understanding of compact objects \cite{VIRGO:2014yos, LIGOScientific:2016aoc, LIGOScientific:2016dsl, LIGOScientific:2020ibl, LIGOScientific:2020kqk, LIGOScientific:2021djp}. The LIGO and Virgo experiments have so far detected around {90} binary compact object mergers composed of black holes (BHs) and neutron stars (NSs) \cite{LIGOScientific:2021djp}, with a few events challenging the models of stellar evolution. LIGO/Virgo's third observing run \cite{LIGOScientific:2020ibl} announced the most massive binary BH (BBH) merger (GW190521) \cite{LIGOScientific:2020iuh} with the primary component mass of $\sim 85\,M_{\odot}$. This observation challenges the formation and evolution mechanisms of stellar BHs since models predict no objects larger than about $65\,M_{\odot}$ due to the pulsational pair-instability (PI) process \cite{Heger:2002by}. This sets a mass gap for BHs between $70 - 150\,M_\odot$ \cite{Belczynski:2016jno}. Primordial black holes (PBHs) formed from the collapse of large density fluctuations in the early Universe \cite{Zeldovich:1967lct, 1971MNRAS.152...75H, 1974MNRAS.168..399C, Carr:1975qj}, could provide an alternative origin for such events.
Another challenge is to model detected pairs which present a large asymmetry in the progenitors masses (\eg GW190412), with mass ratios of nearly 4--to--1 ($30\,M_{\odot} + 8\,M_{\odot}$) \cite{LIGOScientific:2020stg}. A scenario where more than one species of stars or BHs is confined to a small volume could be explained by a sequence of mergers.

Several mechanisms for the origin of the progenitors in BBHs have been considered in the literature; for example, 1)~binary star evolution through common envelope \cite{1973NInfo..27...70T, Bethe:1998bn, Belczynski:2001uc}, 2)~dynamical process in triples \cite{Arca-Sedda:2018qgq}, 3)~pairing of PBHs \cite{Carr:2016drx, Ali-Haimoud:2017rtz}, or 4)~as a product of hierarchical mergers \cite{Fishbach:2017dwv, Gerosa:2017kvu, Liu:2019rnx, Doctor:2019ruh, Wu:2020drm, Kimball:2020qyd, Gerosa:2021mno}. 
In this article we propose a combination of two mechanisms (3 \& 4); \ie PBHs of a common mass pair to form a binary and the ``product" of their merger will pair with another (P)BH, and so on. We explore the possibility of repetition of this process in a ``hierarchical merger" sequence throughout the age of the Universe, considering the radiation of GWs, as they could potentially provide a detectable population of binaries with one or both component masses in the mass gap of stellar BHs.

The astrophysical environments suitable for hierarchical mergers, where multiple GW events will be produced, require a large escape velocity so that merger remnants are efficiently retained. A compact (dense) region will prevent merger remnants from being ejected by GW recoils. Scenarios previously considered for the efficient production of hierarchical mergers include globular clusters \cite{Rodriguez:2016kxx, Askar:2016jwt, Samsing:2017xmd, Mapelli:2021syv}, nuclear star clusters \cite{Kimball:2020opk, Fragione:2020nib, Kritos:2020wcl, Fragione:2021nhb}, and accretion disks surrounding active galactic nuclei \cite{Yang:2019cbr, Tagawa:2020jnc, Tagawa:2020qll}. Here we propose a hierarchical merger of PBHs in Dwarf Galaxies (DGs) for two reasons: First, DGs are dominated by DM \cite{2019ARA&A..57..375S}, which we assume to be in the form of PBHs\footnote{PBHs with a mass larger than $10^{15}$ g are candidates for Dark Matter (DM) \cite{Hawking:1974sw}.}, and second, these environments are dense enough for the hierarchical merger of PBHs \cite{2019ARA&A..57..375S}.

We explore the possible formation of systems like GW190412 and GW190521 as a result of our model (hierarchical merger is also considered in \cite{Rodriguez:2020viw}). We also investigate the formation of BHs in the mass gap and also merged pairs with a large mass ratio \cite{Gerosa:2019zmo}.

The paper is organized as follows: in section~\ref{sec2} we discuss DGs as hosts of PBHs. We also summarize the basic equations of PBH as a candidate for DM \cite{Carr:2020gox}. In section~\ref{sec3}, we explain our model and in section~\ref{sec4} we present our results. Finally, section~\ref{sec5} is devoted to our conclusions.

\section{Dwarf Galaxies and Primordial Black Holes}\label{sec2}
\paragraph{}
Dwarf Galaxies (DGs) are the oldest, and least chemically evolved stellar systems known \cite{2019ARA&A..57..375S}. These galaxies resemble globular clusters in density profiles, however, they are distinguished from star clusters since they present a dynamical mass that is substantially larger than the mass inferred from the luminous stellar population \cite{DES:2015zwj}. The stellar kinematics of DGs confirm that they contain a dominant component of DM which is in contrast with the star clusters. Altogether, these properties indicate that DGs are the astrophysical systems where DM is most abundant.

Since these galaxies are highly DM dominated, with mass-to-light ratios $\sim 1000\,M_{\odot}/L_{\odot}$ \cite{Simon:2007dq}, they are extremely valuable laboratories for characterizing the nature of DM, mostly as probes of DM on the smaller end of the large scale structure spectrum \cite{2018MNRAS.481.5073E}.

Dozens of DGs have recently been discovered by several surveys like the SDSS \cite{Willman:2004kk, Martin:2007ic, 2018PASJ...70S..18H} as satellites of the Milky Way, Andromeda, and as members of the Local Group \cite{Laevens:2015kla}\footnote{Recall that DGs have been a source of a tension known as missing satellites problem since predictions for the abundance of massive satellite galaxies in simulations present significantly larger numbers than the observed objects \cite{1999ApJ...522...82K, 1999ApJ...524L..19M}.}. The ultra-faint Milky Way satellites have masses ranging from just over $10^6 M_{\odot}$ (Coma Berenices) up to $2.8\times10^7 M_{\odot}$ (Canes Venatici I) \cite{Simon:2007dq}. Discovered DGs have Plummer (half-light) radii as small as $\sim 20$ pc \cite{1911MNRAS..71..460P}, with the total mass inside this radius in the range of one to three orders of magnitude larger than their stellar masses \cite{2012AJ....144....4M}.
Consequently, the measured central densities, and the density profiles are essentially those of DM halo. The central densities of DGs range from $\sim 0.08\,M_{\odot}$/pc$^{3}$ up to $\sim 2.1\, M_{\odot}$/pc$^{3}$ \cite{Simon:2007dq}.

The identity of the dominant DM in these environments, however, remains a mystery \cite{Planck:2018vyg}. One appealing possibility is that the DM consists of black holes formed in the early Universe, known as Primordial Black Holes (PBHs) \cite{Carr:2016drx, Green:2020jor}. Constraints on their abundance imposed in DG environments are reported in Refs.~\cite{Zhu:2017plg, Stegmann:2019wyz}\footnote{Recently, some studies examined whether DM candidates in the form of PBHs can solve the cusp-core problem in low-mass DGs \cite{Boldrini:2019isx}.}. In the following, we will briefly explain the properties of PBHs and their abundances.

The most common mechanism for PBHs formation is the collapse of density fluctuations larger than a threshold that reenters the horizon after inflation in the early Universe \cite{Zeldovich:1967lct, 1971MNRAS.152...75H, 1974MNRAS.168..399C}\footnote{See however \cite{Padilla:2021zgm} for an alternative, more likely mechanism during a long-lasting period of reheating.}. Thus the mass of PBH is roughly the mass of the horizon, $M_{\rm H}$
\be\label{mass}
M_{\bh} \approx M_{\rm H} \sim 10^{15} \lp\dfrac{t}{10^{-23}\,{\rm s}}\rp {\rm g}\,.
\ee
PBHs radiate thermally due to the Hawking radiation \cite{Hawking:1974sw}, but massive enough PBHs survive for more than a Hubble time. In consequence, their mass spectrum spans many orders of magnitude, from $\sim 10^{15}$ g to well over $10^{50}$ g.

A parameter which represents the abundance (the energy density fraction) of PBHs in the epoch of their formation is
\be\label{beta1}
\beta \equiv \frac{\rho_{\bh}(t)}{\rho(t)}\,.
\ee
Assuming an adiabatic expansion of the Universe, the above can be related to the present abundance mass fraction of PBHs, $f_{\bh}\equiv\Omega_{\bh}/\Omega_{\rm DM}$, by \cite{Carr:2020gox}
\be\label{beta2}
\beta \simeq 3.7\times10^{-9}\,\lp\frac{g_*}{10.75}\rp^{1/4}\lp\frac{M_{\bh}}{M_{\odot}}\rp^{1/2} \,f_{\bh}\,,
\ee
where $g_{*}$ is a number of relativistic degree of freedom at the time of PBHs' formation.

A population of PBHs with a specific mass is subject to observational constraints which in many cases reduces the contribution of $f_{\rm PBH}$ to Dark Matter. However, DM can be completely formed of PBHs in the mass window around $(5\times10^{-16}-2\times10^{-14})\,M_{\odot}$  \cite{Carr:2020gox, Franciolini:2022htd}. The recent detection of GWs by LIGO/Virgo also imposes constraints on the abundance of PBHs with masses of order $10$ to $50\,M_{\odot}$. Searches for compact objects of sub-Solar masses also impose constraints on PBH abundance \cite{Authors:2019qbw}. There are also constraints on sub-Solar masses by gravitational microlensing observations \cite{Niikura:2017zjd}. These constraints are imposed without any assumption about the distribution of PBHs at their formation, their clustering, or the environment, which can alter the bounds significantly \cite{Clesse:2016vqa, Belotsky:2018wph}. In the next section, in light of the observational constraints, we will consider DM PBHs with four different masses.

\section{Model for the Merger Rate}\label{sec3}
\paragraph{}
In what follows we describe the details of the model employed to produce a succession of mergers of BHs in a DG. We consider a population of PBHs at the core of the DG, with $N_{\bh}$ elements, which we call a {\it cluster}. We assume PBHs are the only component of DM, with a single initial mass (monochromatic population). We test four different initial masses of PBHs, where $N_{\bh}$ is determined by the mass of DG. Throughout this work we consider a DG with the total mass, $M_{\rm DG}=10^{9}\,M_{\odot}$, and radius, $R_{\rm DG}\sim 10$ pc.

\begin{figure*}[h!]
\centering
\includegraphics[width=14cm, height=11cm]{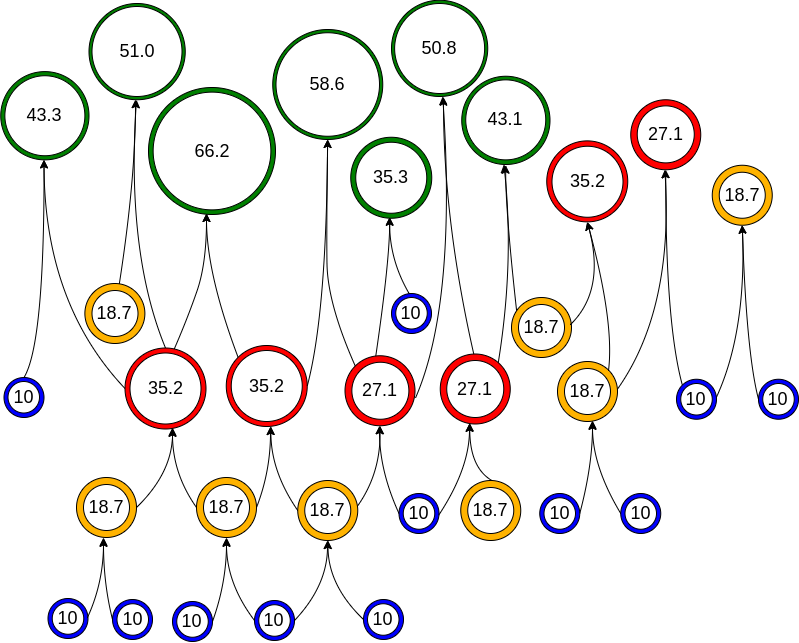}
\caption{The hierarchical tree of (P)BH mergers from an initial population of PBHs with mass $10\,M_{\odot}$. In the age of the Universe, collisions reach up to a fourth generation of BHs. Different colors indicate different generations, and approximate masses, in units of $M_{\odot}$, consider the mass-loss due to GW radiation.}
\label{tree}
\end{figure*} 

In order for PBHs to merge more than once, a dense environment is required. We thus focus on core of DGs with mass, $M_{\rm c} = 10^5\,M_{\odot}$ confined in a core radius, $R_{\rm c} = 0.9$ pc, which is admitted by observations \cite{Simon:2007dq}. The number of PBHs is proportional to the density of DM, $\rho_{\rm DM}$,
\be\label{number}
N_{\bh} = \dfrac{4\pi}{3}\dfrac{\rho_{\rm DM}\,R_{\rm c}^3}{m_{\bh}}\,.
\ee

In our analysis we consider four different initial masses for PBHs, which we define as {\it first generation} (1G); $10^{-14}\,M_{\odot},\,10^{-2}\,M_{\odot},\,1\,M_{\odot}$ and 10 $M_{\odot}$. We choose $10^{-14}\,M_{\odot}$ since as mentioned there is no confirmed observational constraints for this mass and PBHs can be whole DM. Our interest in other mass values comes from the GWs observations. From Eq.~\eqref{number}, we get the number of PBHs at the core (see Table~\ref{table} below).

According to our assumption, the components of the first merger will have equal masses. Therefore, the merger of 1G+1G will produce the second generation BHs (2G) with roughly double the mass. Subsequently, from the third generation (3G) onward, BHs can form from 1G+2G and 2G+2G; two populations of BHs will form, and therefore, a variety of masses is possible (see Figure~\ref{tree}). 

For estimating the merger rate, we know that the cross-section of two objects with masses $m_{i}$ and $m_{j}$ which merge after first being captured into binaries by the emission of gravitational radiation is \cite{1989ApJ...343..725Q, Mouri:2002mc}
\be\label{crosssection}
\sigma= \dfrac{\sigma(m_i,m_j)}{|v_{i}-v_{j}|^{18/7}}\,,
\ee
with 
\be\label{sigmaij}
\sigma(m_i,m_j) = 2 \pi \left(\frac{85\pi}{6\sqrt{2}}\right)^{2/7}\frac{G^{2}(m_{i} +  m_{j})^{10/7}\lp m_{i}m_{j}\rp^{2/7}}{c^{10/7}}\,.
\ee

\noindent Here $v_{i}$ represents the velocity of the components in the binary, which is given by the root mean-squared (rms) velocity, $\overline{v_i} \equiv \sqrt{\overline{v_i^2}}$ 
\be\label{velocity}
\overline{v_{i}^{2}}(r) = \frac{4\pi}{n_{i}(r)}\int_{0}^{\phi(r)}\,f_{i}(E) \lb2\lp\phi(r)-E\rp\rb^{3/2}\,dE\,,
\ee
where $n_{i}(r) = \dfrac{\rho_{i}(r)}{m_{i}}$ is the number density, and the DM density is modelled by a Plummer sphere \cite{1911MNRAS..71..460P} at all times. This prescribes the following expressions for the density and the gravitational potential, respectively \cite{1911MNRAS..71..460P}
\bea
\rho_{i}(r) &=& {\rho}_i^{\rm c}\lp 1 + \frac{r^{2}}{R_{\rm c}^{2}}\rp^{-5/2} = \frac{3\,m_{i}\,N_{i}}{4\pi\,R_{\rm c}^{3}}\lp 1 + \frac{r^{2}}{R_{\rm c}^{2}}\rp^{-5/2}\,,\\
\label{gravpot}
\phi(r) &=& {\phi}^{\rm c}\lp 1 + \dfrac{r^{2}}{R_{\rm c}^{2}}\rp^{-1/2} = \dfrac{GM_{\rm c}}{R_{\rm c}}\lp 1 + \dfrac{r^{2}}{R_{\rm c}^{2}}\rp^{-1/2}\,,
\eea
where we defined the core density of the $i$-th species, ${\rho}_i^{\rm c} \equiv \frac{3\,m_{i}\,N_{i}}{4\pi\,R_{\rm c}^{3}}$, with  $N_i$ representing the total number of the BHs with mass $m_i$, as well as a \textit{core potential}, ${\phi}^{\rm c} \equiv \frac{GM_{\rm c}}{R_{\rm c}}$. The distribution function of each BH population, $f_i(E)$ is given by \cite{1989ApJ...343..725Q}
\be\label{DF}
f_{i}(E) = \frac{32\sqrt{2}}{7\pi^2}\,(\phi^{c})^{-5}\,n_i^c\,E_i^{7/2}\,,
\ee
where $n_i^c \equiv n_i(r=R_c)$, and according to the virial theorem, the energy of each species of the {\it cluster} is given by
\be\label{energy}
E_{i} = \dfrac{G\,m_{i}^2\,N_{i}^{2}}{2\,R_{\rm c}}\,.
\ee
An approximate analytic expression for the merger rate in this case is estimated from the Fokker-Planck equation, in particular, from the terms that account for the loss and gain of BHs due to mergers with other BHs \cite{1989ApJ...343..725Q}  (see also \cite{Stasenko:2021vmm}). Once the velocities take the rms value, this can be expressed as
\be\label{rate}
\Gamma_{j} = \frac{14\pi}{3} \sum_{i}\sigma(m_{i},\,m_{j})\int\;dr\,r^{2}\,\frac{n_{i}}{\overline{v}_{i}}\frac{n_{j}}{\overline{v}_{j}} \lb\lp\overline{v}_{i} + \overline{v}_{j}\rp^{3/7} - |\overline{v}_{i} - \overline{v}_{j}|^{3/7}\rb\,.
\ee
The rms velocity is the solution of Eq.~\eqref{velocity}, which in light of the above definitions, can be expressed as
\be
\overline{v_{i}}^2(r) = \frac{1}{2}\phi(r) \lp\frac{\phi(r)}{\phi^c}\rp^5 \lp \frac{n_i(r)}{n^c_i}\rp^{-1}\,.
\ee

In order to define the merging components, we take as different elements the  species, $i$ or $j$, defined by their masses, and we also divide the galaxy core in shells which components will collide with a distance-dependent merger rate. From the expression in Eq.~\eqref{rate}, the merger rate in region between shells at radii $r_A$ and $r_B$ is given by
\be
\begin{split}
\Gamma&(r_A,\,r_B) = \frac{14\pi}{3} \sum_{i}\sigma(m_{i},\,m_{j})\,\frac{n_j(r_{B}) n_i^c}{\overline{v_{j}}(r_{B})}\,\lp\frac{\phi^c}{2}\rp^{-2/7} \int\,dr_{A} \,r^{2}_{A}\,(1+A)^{-9/4}\\
& \times \left\{\left[\left(\frac{1}{(1+B)^{1/4}}\right) + \left(\frac{1}{(1+A)^{1/4}}\right) \right]^{3/7} - \left|\left(\frac{1}{(1+B)^{1/4}}\right) - \left(\frac{1}{(1+A)^{1/4}}\right)\right|^{3/7}\right\}\,,
\end{split}
\label{final:gamma}
\ee
where $A=(r_{A}/R_{\rm c})^2$ and $B=(r_{B}/R_{\rm c})^2$, label the shell to which each of the merging components belong. 

For each example, we have sliced the galaxy core in 10 shells and we proceed to count the number of mergers by dividing the cosmic time in \textit{merger epochs} of constant $\Gamma(r_A,\,r_B)$. The lasting time for such epochs, $\tau_e$ is defined as one tenth times the inverse of the largest merging rate value among the pairs of shells. Mathematically, 
\be
\tau_e = \frac{1}{10} \mathrm{min}\left\{\Gamma(r_A,\,r_B)^{-1}\right\}\,.
\ee

We count the number of mergers, starting with a  single population of PBHs at redshift $z = 20$. This initial condition sets an early enough time to discard the stellar origin for the initial BHs, and late enough for the formation of DGs. For the first iteration, the initial single-mass population is divided into two identical sets, taken as two different species as an input of our merging algorithm. In every epoch, we evaluate the probability of encounters from the merger rate in Eq.~\eqref{final:gamma} for a given combination of two species at two given radii ($r_A$ and $r_B$), verifying that the merger time is shorter than the period each merger epoch lasts. We thus count the number of mergers by considering the number of BHs available in each shell for both species. For the next epoch, we reset the merging rates at time $t_{i} + \tau_e$ for the updated population of each species. Thus the above procedure is repeated to compute the mergers at each shell. In Figures~\ref{numbers} and \ref{relative}, we report results for the total number of BHs formed in the whole DG core (after integrating all shells). 

In each merger, we also consider the mass loss due to the emission of gravitational radiation. The binary is formed through gravitational capture where the energy is released through the emission of GW. Such dissipative binary formation is followed by the coalescence of the binary with a well-known rate of energy loss. The time-averaged energy loss rate of the binary in the Keplerian orbit is given by \cite{PhysRev.136.B1224}
\be
\bigg\langle\dfrac{dE}{dt}\bigg\rangle = -\dfrac{32}{5}\dfrac{G^4\,(m_im_j)^2\,(m_i+m_j)}{a^5} F(e)\,,
\ee
where $e$ and $a$ are eccentricity and semi-major axis of the orbit of binary, respectively, and the explicit dependence on the eccentricity is given by
\be
F(e) = \frac{1}{(1-e^2)^{7/2}}\lp 1+\dfrac{73}{24\,}\,e^2 + \dfrac{37}{96}\,e^4\rp\,.
\ee

\noindent Since the orbits follow initially a parabolic path, which is also required by the cross-section expressed in Eq.~\eqref{crosssection}, we take the eccentricity as $e = 0.99$ as an approximation to the parabola and the semi-major axis as the initial (periastron) separation. The energy emission rate can be expressed as a function of the semi-major axis decrease \cite{PhysRev.136.B1224}
\be
\frac{da}{dt} = -\frac{64}{5}\frac{G^{3}}{c^{5}}\frac{m_{i}m_{j}\,(m_{i}+m_{j})}{a^{3}}F(e)\,, 
\ee

\noindent Therefore the energy loss in GWs throughout the merger is given by
\be\label{Eq:mass-loss}
\Delta E = \int_{a_{i}}^{a_{\rm merge}}\frac{dE}{da}\, \frac{da}{dt}\,dt = \frac{1}{2}m_{i}m_{j}\lp\frac{1}{a_{\rm merge}} - \frac{1}{a_i}\rp\,,
\ee
where the semi-major axis is integrated from an initial separation, $a_i$ up to the sum of the Schwarzschild radii of the progenitors, $a_{\rm merge}$. The energy emitted by gravitational radiation is accounted for by the effective mass loss. This results in the merger product with a mass smaller than the sum of the progenitors' masses, as illustrated in Figure~\ref{tree}. The accumulated effect on the total mass of the {\it cluster} is also taken into account in our evaluation of merger rates (see Table~\ref{table}). Our prescription for the mass loss is in agreement with events observed by the LIGO/Virgo collaboration\footnote{The computed mass via Eq.~\eqref{Eq:mass-loss} from the progenitor components deviates in average only 2\% from the final mass estimated for the events in the three runs of LIGO/Virgo (well within the determined errors).}. Finally, it is worth mentioning that we do not consider the disruption of BHs binaries by encounters with other compact objects (BHs and stars). We also neglect the ejection of BHs due to the recoil kicks of GW radiation. Note that even if we take these effects into account, they will only affect an insignificant fraction of the total number of merger products.

\section{Results}\label{sec4}
\paragraph{}
In this section, we present the results of a succession of merger epochs according to the model described above. We consider DG cores dominated by DM PBHs, with four cases of a single initial PBH mass given by $m_{\rm PBH}/M_{\odot} = \lp 10^{-14},\,10^{-2},\,1,\,10\rp$. We keep the initial mass of the core as a constant, $M_{\rm c}(t_i) = 10^5 M_{\odot}$, which implies that the initial number of PBHs and number density is sensitive to the initial mass of (P)BHs (see Table~\ref{table}). For illustration purposes, in Figure~\ref{tree} we show the merger tree of (P)BHs, for the specific case of initial mass, $m_{\rm PBH} = 10\,M_{\odot}$ (blue circles). Each merger gives way to a BH of the $n$-th generation if at least one of its progenitors belongs to the $n-1$-th generation. Thereby, the second generation is constituted by the 1G+1G progenitors (yellow circles), while mergers of 1G+2G and 2G+2G BHs constitute the third generation (red circles), and so forth. In our approach, only four epochs fit within the age of the Universe and thus, BHs merge up to the fourth generation, because the $n$-th epoch will show mergers from the first up to the $n$-th generation. As stated earlier, due to the GW radiation in each merger, the mass of each merger product is smaller than the sum of the components of the binary.  

\begin{table}[h!]
\begin{center}
\begin{tabular}{|c|c|c|c|}
\hline
$m_{\bh}$   &  $N_{\bh}(t_{i})$   &  $M_{\rm c}(t_{0}) / M_{\rm c}(t_{i})$ &  
$\sum m_{\rm PBH}(t_{0})/M_{\rm c}(t_{0})$ \\ \hline  \hline
$10^{-14}\,M_{\odot}$ & $10^{19}$  & $0.953$ & $0.5897$ \\ \hline
$10^{-2}\,M_{\odot}$  & $10^{7}$   & $0.952$ & $0.5897$ \\ \hline
$1\,M_{\odot}$        & $10^{5}$   & $0.952$ & $0.5896$ \\ \hline
$10\,M_{\odot}$       & $10^{4}$   & $0.953$ & $0.5891$ \\ \hline
\end{tabular}
\caption{The first and second columns show the mass and number of the original  population of PBHs (1G), respectively. The third column represents the remaining mass of the core after four epochs. The final column shows the final (present) percentage of the original PBH population in terms of mass.}
\label{table}
\end{center}
\end{table}
Table~\ref{table} indicates the number of PBHs in the initial population (at time $t_i$) and the remaining mass of the core after the four merger epochs  at the present time, $t_0$, ($M_{\rm c}(t_0) / M_{\rm c}(t_i)$)--third column. The last column shows the mass fraction of the initial PBH population which did not collide with another BH up to $t_0$. It is worth noting that these results are practically independent of the initial mass of PBHs. 

\begin{figure*}[h!]
\centering
\subfloat[]{\includegraphics[scale=0.5]{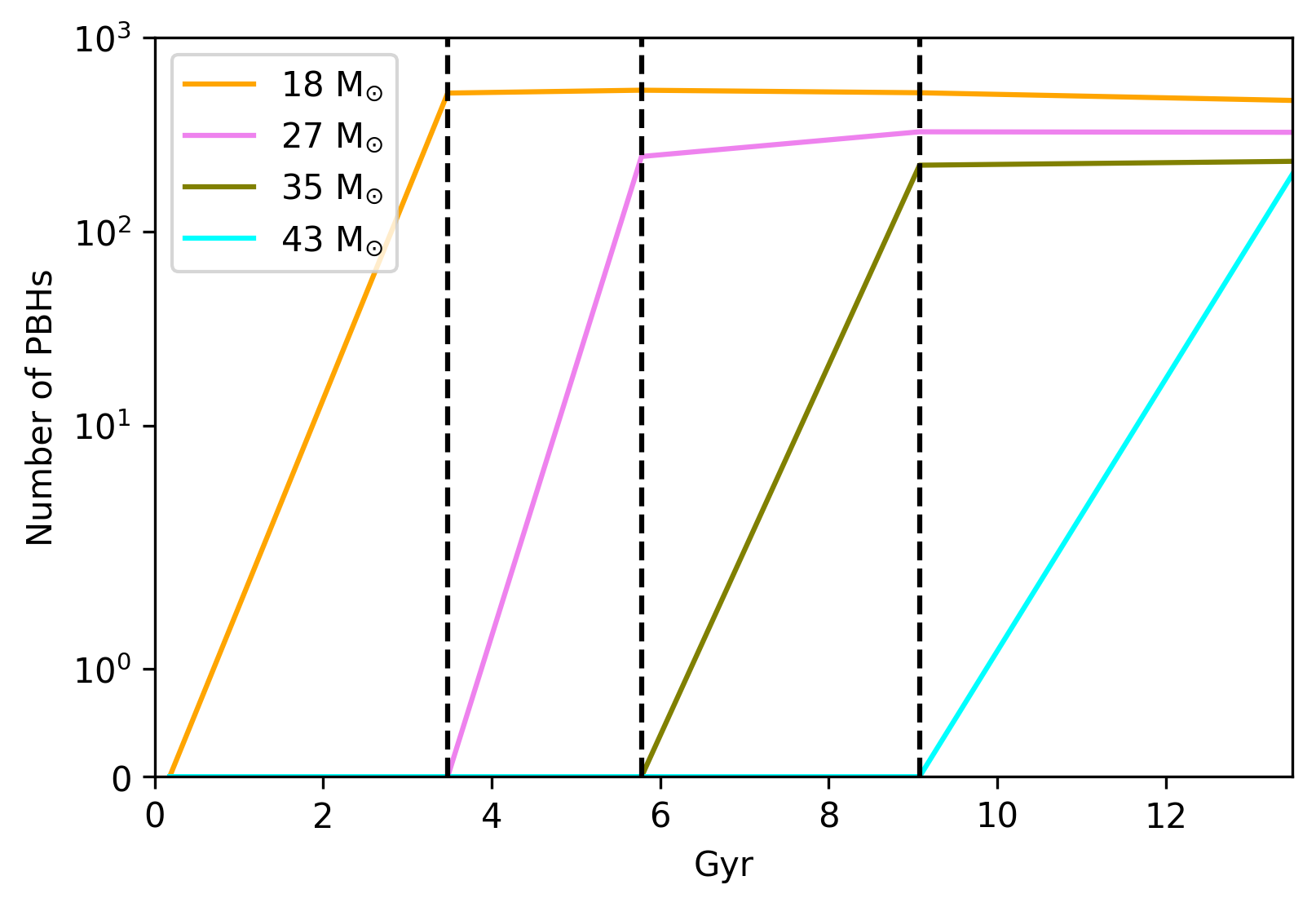}}
\qquad
\subfloat[]{\includegraphics[scale=0.5]{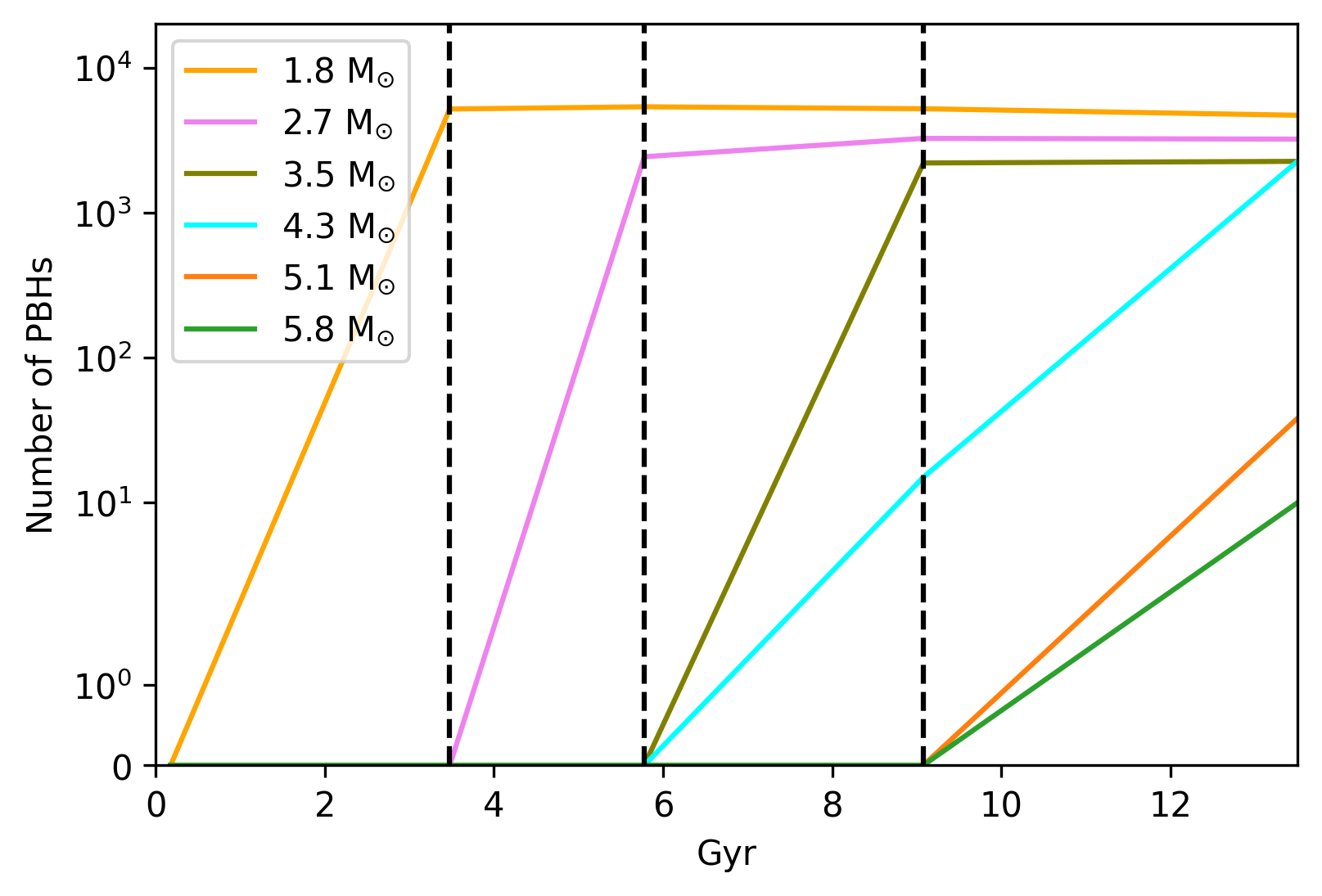}}
\quad
\subfloat[]{\includegraphics[scale=0.5]{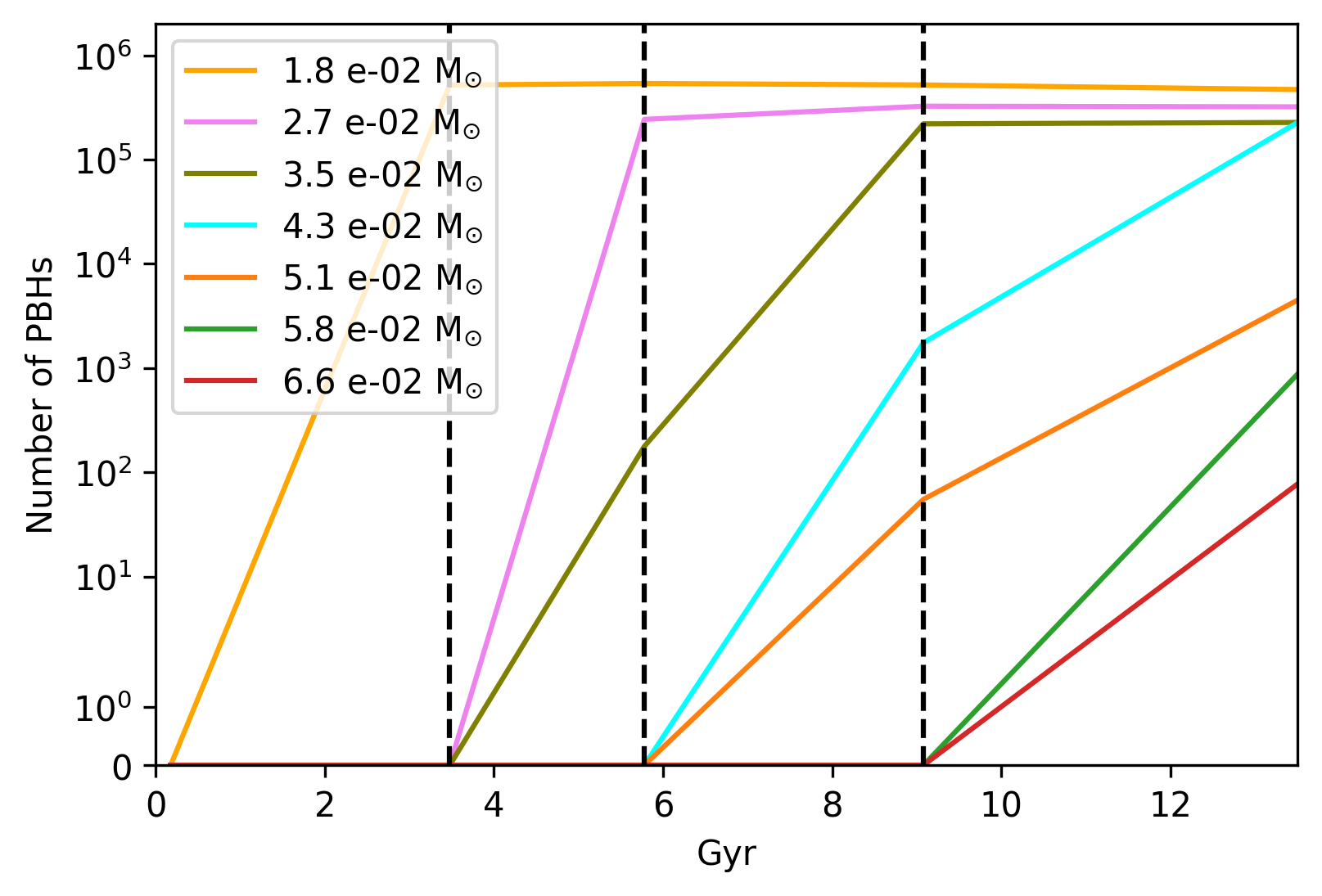}}
\qquad
\subfloat[]{\includegraphics[scale=0.5]{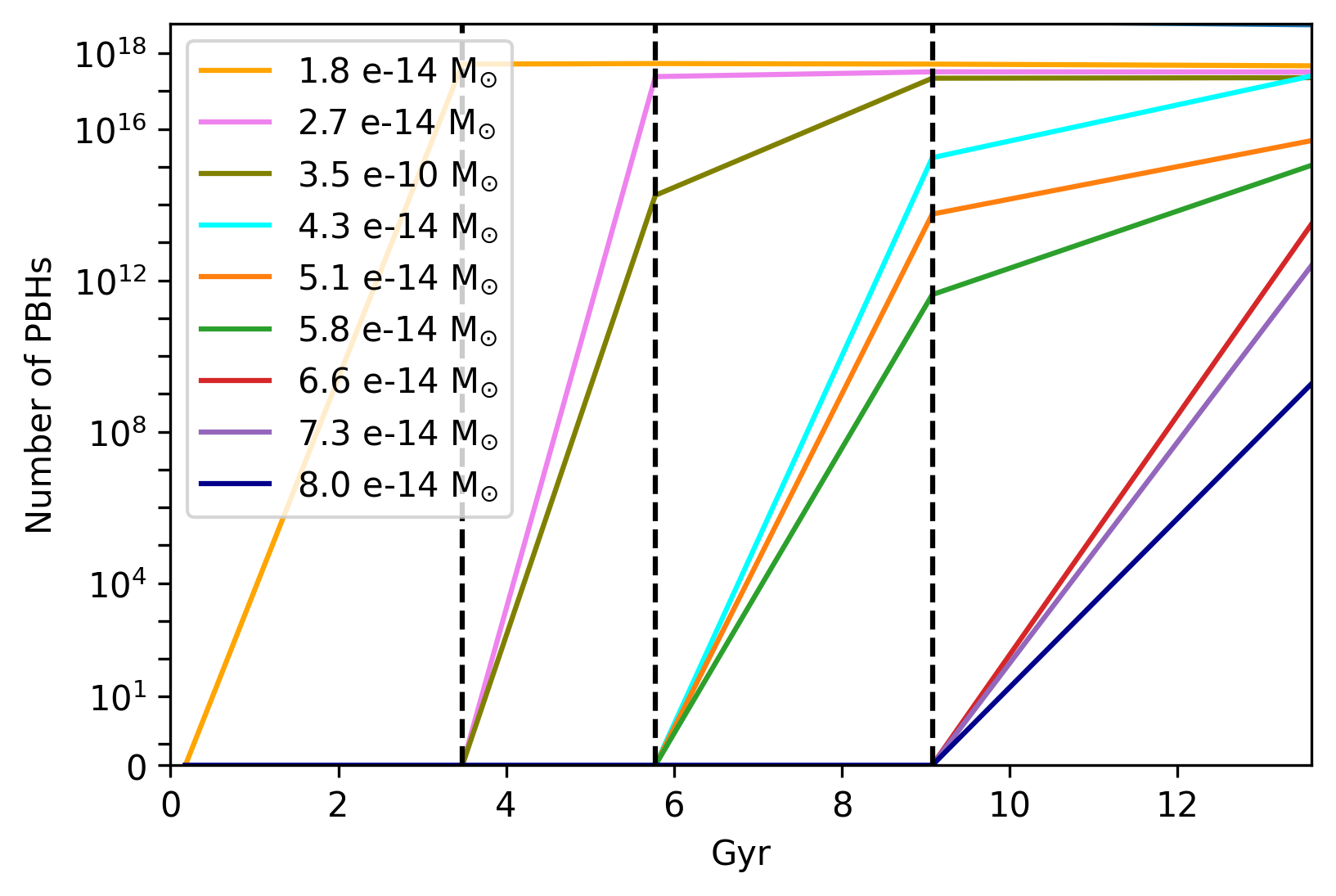}}
\caption{Number of BHs formed from mergers after each epoch, shown as a function of the age of the Universe, starting from $t = 0.18~\mathrm{Gyr}$ ($z = 20$). Each epoch is marked with a vertical dashed line. The panels correspond to the following masses for the initial populations: (a) $10\,M_{\odot}$, (b) $1\,M_{\odot}$, (c) $10^{-2}\,M_{\odot}$, (d) $10^{-14}\,M_{\odot}$.}
\label{numbers}
\end{figure*}

Figure~\ref{numbers} shows the number count of merger products at each epoch for the different initial populations of PBHs considered. Note that for smaller PBH masses, merger products of mass up to $8\, m_{\rm PBH}$ are possible. On the other hand, for large initial PBH masses, only BHs of mass $4.3\, m_{\rm PBH}$ are significantly produced. This is due to the number of PBHs initially present, as listed in Table~\ref{table}. Note also that in Figure~\ref{numbers}, the dominant product populations in numbers are those of the first and second generation of mergers. This is true for all investigated cases and is due to the fact that the relative merger rates are independent of the PBH mass but depend on the number of PBHs.   

\begin{figure*}[h!]
\centering
\subfloat[]{\includegraphics[scale=0.5]{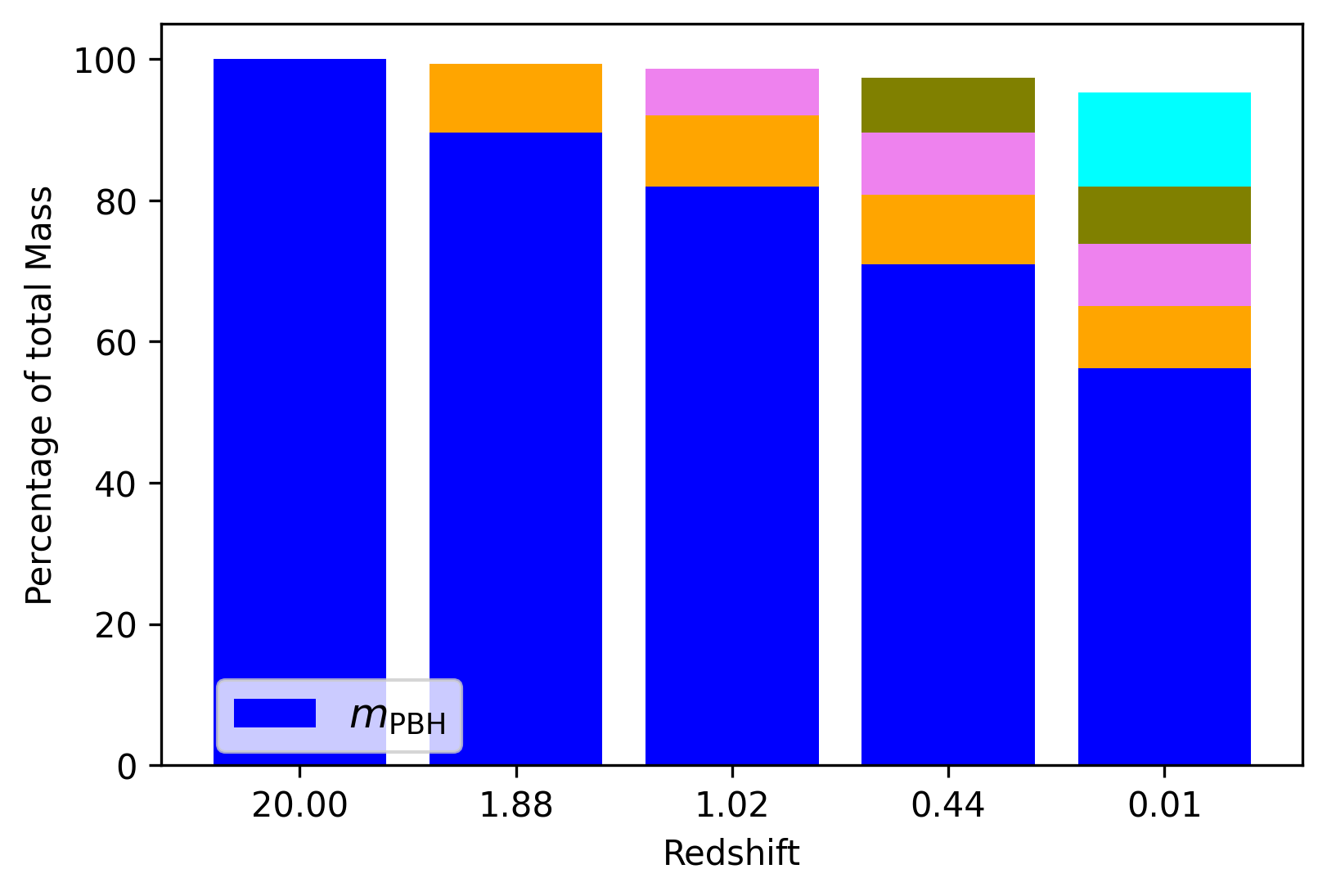}}
\qquad
\subfloat[]{\includegraphics[scale=0.5]{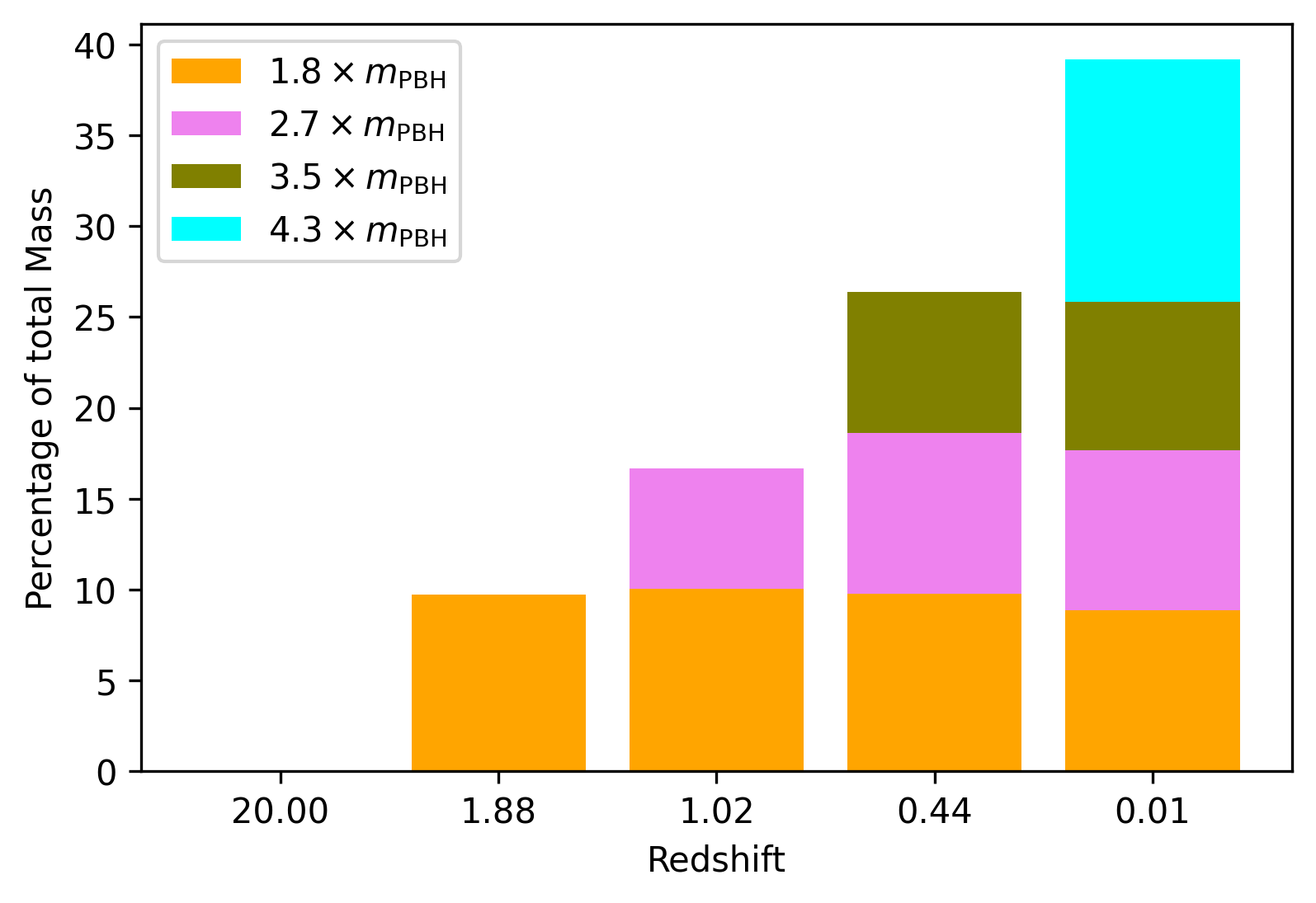}}
\caption{Percentage of the total mass in each population (of a given mass) for the most abundant populations. This is plotted for the case where $m_{\rm PBH} = 10\,M_{\odot}$ but the fractions are largely independent of the PBH mass. The original population is included in Panel (a), where part of the missing mass is in smaller species, and only $4.7\%$ is lost in the radiation of gravitational waves. As shown in Panel (b), the percentage of mass in the three most numerous populations generated is of order $10\%$ each, while the most massive population constitutes more than $15\%$ of the final mass. Note that these percentages are the same for a range of initial masses of PBHs in the {\it cluster}. Finally, note that the plots account only for the mass contribution of each population, and therefore we make no distinction among the generation in which they are formed.} 
\label{relative}
\end{figure*}

Focusing on the galaxy core mass fraction of the merger products, we plot their contribution in Figure~\ref{relative}, where we show the relative mass abundance of each population in a DG core initially formed of PBHs with mass $m_{\rm PBH}$. We find that the resulting mass fractions are independent of the value of the original PBH mass (see Table~\ref{table}). The left plot of this figure shows that the fraction of PBHs decreases steadily with time, increasing the mass fraction of merger products after each merger epoch. After the four iterations, we find that more than 40\% of the PBHs have formed at least one binary and merged to form larger BHs. The right panel of Figure~\ref{relative} shows the mass fraction in merger products. Note that the mass fraction is dominated by the more massive species even when the number count of such species is subdominant (see the cyan component in Figures~\ref{numbers}~and~\ref{relative}).\\

\section{Summary and Discussion}\label{sec5}
\paragraph{}
Primordial black holes formed in the early Universe, before the formation of stars, can exist as dark matter and also contribute to the black hole merger events observed through gravitational waves. GW observations have demonstrated that BHs mergers may be more frequent than expected. If merger products form new binaries, they may subsequently merge as detectable GW sources. For sequential mergers of BHs, a dense, DM-dominated environment is required. Dwarf galaxies are ideal scenarios to host a hierarchical merger of BHs. 

In this paper, we studied sequence mergers of BH starting with a monochromatic {\it cluster} of PBHs at the core of DGs. Our study, featuring stages of constant merger rate, represents a first step towards the full numerical analysis of the Fokker-Planck equation. Since PBHs span several decades of mass, we considered four different initial masses. Starting the evolution of the system at redshift $z = 20$, we find that in cores with massive PBHs ($\sim 10\,M_\odot$), BHs with up to four times the initial mass can be formed significantly, while in the small PBH mass limit the masses of products are up to eight times the initial PBH mass. Our results also show that the total mass loss at the DG core from GW emission, and the mass fraction of BH that undergo collisions are mostly independent of the initial PBH mass. Figure~\ref{relative} in particular shows that the original population of PBHs is reduced to little more than 50\% of the total mass. The proportion of larger mass BH populations may be tested by future GW surveys.
 
Our results are not only relevant to the rate and spectrum of GW events. We have shown that binary formation in the dense cores of DM-dominated systems can give way to more than one population of PBHs. Our results are also in agreement with studies investigating systems of multiple stars/BHs in the central regions of clusters with an accretion disk \cite{McKernan:2014oxa, Miralda-Escude:2000kqv}, and without it \cite{Mapelli:2021syv, Rodriguez:2020viw, Martinelli:2022elq}.   
There are important differences in the populations of BH binaries that may distinguish our merger scenario (see \eg \cite{Gerosa:2017kvu}). 
Ultimately, it is important to assess if the stochastic GWs of the proposed mechanism yield a detectable signal (see \eg \cite{Garcia-Bellido:2021jlq}). There is, however, room for important complements to our assumptions before producing accurate forecasts. For example, since the DGs are faint with low stellar mass, we have neglected the effect of mergers on stars, but a more detailed study contemplating such collisions would include electromagnetic signals which will constrain the parameters of the model.
The recoil velocity of the product BH after a merger event is also an ingredient to be included in future studies \cite{Fitchett:1983, Gonzalez:2006md, Varma:2022pld}. In the meantime, our results indicate that a series of sequential mergers may take place at the cores of DGs. 

\section*{Acknowledgments}
We are grateful to Luis Padilla for his insightful comments on the manuscript.
EE thanks the TWAS 2021 Fellowship for Research and Advanced Training. This work is sponsored by CONACyT grant CB-2016-282569 and by Program UNAM-PAPIIT Grant IN107521 ``Sector Oscuro y Agujeros Negros Primordiales''.

\end{document}